\documentclass[10pt]{article}
\usepackage[utf8]{inputenc}
\usepackage[T1]{fontenc}
\usepackage[numbers,sort&compress]{natbib}
\usepackage{float}
\usepackage{authblk}
\usepackage{amsmath}
\usepackage{graphicx}
\usepackage{hyperref}
\usepackage[margin=1in]{geometry}
  
\title{\textbf{Statistical Methods for Determining Turbulence in Supercontinuum Generation}}
 
\author[1]{Mehmet Müftüoglu}
\author[1,2,*]{Mario Chemnitz}
 
\affil[1]{Leibniz-Institute of Photonic Technology, Albert-Einstein-Str.~9, 07745 Jena, Germany}
\affil[2]{Institute of Applied Optics and Biophysics, Friedrich Schiller University Jena, Albert-Einstein-Str.~15, 07745 Jena, Germany}
\affil[*]{Corresponding author: \texttt{mario.chemnitz@leibniz-ipht.de}}
 
\date{}

\begin{document}

\maketitle
\thispagestyle{empty}

\begin{abstract}
Distinguishing coherent, turbulent, and chaotic operating regimes in supercontinuum generation is important for understanding nonlinear optical dynamics and optimizing broadband light sources. Experimentally identifying the onset of turbulence remains challenging because the most common metric, first-order coherence, requires access to the complex optical field and cannot be directly obtained from intensity-only measurements. In this work, we investigate whether experimentally accessible statistical observables can identify turbulence in supercontinuum generation. We compare wavelength-integrated variance and kurtosis with simulation-based first-order coherence over a chirp-controlled pulse-duration sweep implemented through additional $\beta_2$ dispersion. The study combines generalized nonlinear Schrödinger equation simulations with shot-to-shot dispersive Fourier transform measurements validated against optical spectrum analyzer spectra. Statistical intensity distributions were analyzed using histograms, complementary cumulative distribution functions, and kurtosis measurements across the generated supercontinuum bandwidth. Simulations and experiments both revealed heavy-tailed intensity statistics in the intermediate pulse-duration regime associated with reduced spectral coherence. The integrated kurtosis reached a maximum near 600 fs in simulations and near 700 fs in experiments, while the integrated variance within the first 20 dB spectral range decreased with increasing pulse duration. The agreement between simulations and experiments demonstrates that variance- and kurtosis-based observables can serve as experimentally accessible indicators of turbulence in supercontinuum generation. These results show that intensity-only statistical measurements can distinguish coherent and incoherent operating regimes without requiring direct field-resolved coherence measurements.
\end{abstract}

\flushbottom
\maketitle

\section*{Introduction}
Supercontinuum generation in nonlinear optical fibers provides a rich platform for studying complex nonlinear wave dynamics. The physical origins of these dynamics lie in third-order nonlinear wave interactions, which give rise to a number of phenomena. Specifically in anomalous dispersive waveguides, the subtle interplay between higher-order soliton fission and modulation instability plays a central role in the emergence of nonlinear coherent structures, breathers, and stochastic wave interactions in dispersive nonlinear systems~\cite{zakharov_modulation_2009}. 

In optical fibers, modulation instability can evolve into highly dynamic regimes involving Akhmediev breathers, soliton fission, and rogue-wave formation~\cite{akhmediev_extreme_2009,hammani_spectral_2011,dudley_instabilities_2014}. The observation of optical rogue waves during supercontinuum generation demonstrated that rare and extreme fluctuations can emerge from noise-seeded nonlinear propagation~\cite{solli_optical_2007,dudley_harnessing_2008}. Subsequent studies further showed that these fluctuations exhibit strong wavelength dependence and heavy-tailed statistical behavior associated with collision dynamics and incoherent spectral evolution~\cite{erkintalo_statistical_2010,borlaug_extreme_2008,kraych_statistical_2019}.
The development of real-time dispersive Fourier transform (DFT) techniques enabled shot-to-shot observation of spectral fluctuations during supercontinuum generation and modulation instability~\cite{wetzel_real-time_2012}. 

Using these measures, several statistical indicators have been introduced to characterize coherent and turbulent nonlinear regimes, including entropy-based analysis, statistical sampling, soliton number scaling, Pearson correlation analysis, and first-order spectral coherence~\cite{dudley_instabilities_2014,perego_complexity_2022}. In particular, first-order coherence has become an important benchmark metric because it directly quantifies shot-to-shot phase stability of the generated optical field. Pearson correlation analysis has also been widely used to identify modulation-instability structures and spectral coupling dynamics. Alternatively, machine-learning-based analysis has been proposed to infer hidden temporal dynamics from experimentally measured spectral fluctuations~\cite{narhi_machine_2018}.

Despite these advances, important experimental limitations remain. First-order coherence requires access to the full complex optical field, whereas DFT measurements provide intensity-only information and therefore do not directly preserve the field spectral phase. Similarly, Pearson correlation analysis identifies spectral relationships but does not directly characterize the probability distribution of the sampled fluctuations or quantify the occurrence probability of rare extreme events. In contrast, statistical observables such as variance, kurtosis, and complementary cumulative distribution functions (CCDFs) can be directly extracted from experimentally measured intensity distributions. These observables provide information about heavy-tailed statistics, intermittency, and extreme-event probability without requiring field measurements. In particular, kurtosis has recently emerged as a useful indicator for identifying nonlinear complexity and modulation-instability-driven fluctuations in optical systems~\cite{perego_complexity_2022}.

In this work, we investigate experimentally accessible statistical observables to identify the onset of turbulence in supercontinuum generation as a function of pulse chirp. We combine generalized nonlinear Schrödinger equation simulations with shot-to-shot DFT measurements validated against optical spectrum analyzer measurements. By comparing integrated variance and kurtosis with simulation-based first-order coherence over a chirp-controlled pulse-duration sweep, we evaluate the extent to which intensity-only statistical metrics can distinguish coherent, turbulent, and incoherent supercontinuum regimes. Finally, we discuss the practical use cases of our approach in the context of nonlinear systems control and potential applications.

\section*{Results}

\subsection*{DFT Validation And Operating Conditions}
\begin{figure}[H]
    \centering
    \includegraphics[width=14 cm]{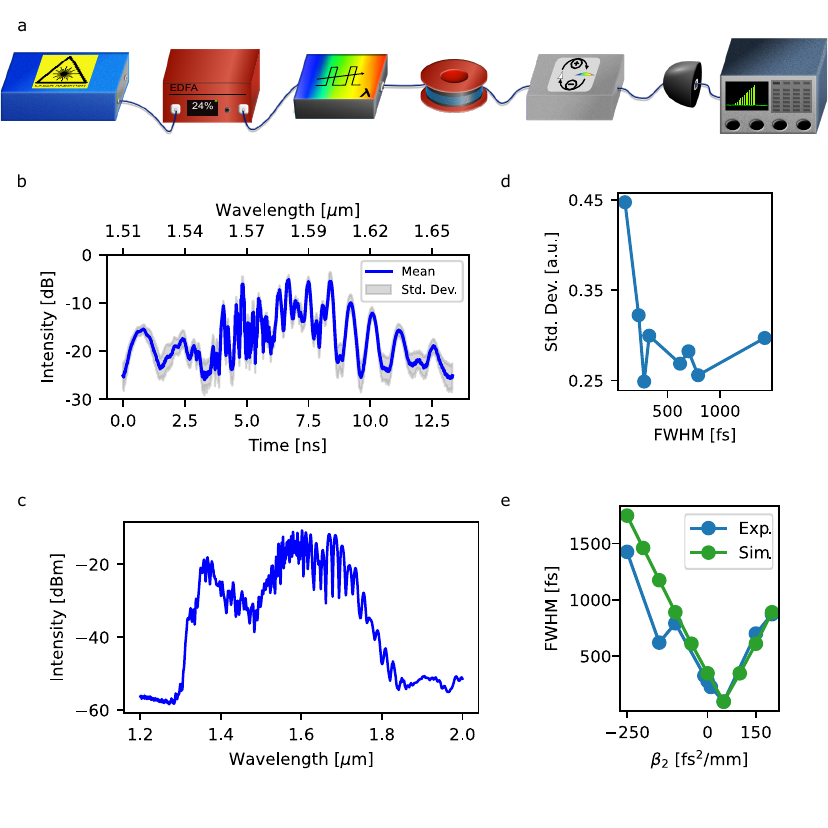}
    \caption{Shot-to-shot ultrafast optical pulse spectrum sampling scheme. a) Sketch of experimental setup for shot to shot spectrum sampling. The experimental setup consists of ultrafast femtosecond pulsed laser, Erdium-doped fiber amplifier (EDFA), spectral wave encoder (WaveShaper), highly nonlinear fiber (HNLF), dispersion compansated fiber (DCF), ultrafast photodiode (UPD), and real-time oscilloscope (RTO). b) RTO sampled spectrum. c) Optical Spectrum Analyzer (OSA) sampled spectrum. d) Chirped optical pulse vs integrated standard deviation (Std. Dev.). e) Chirp vs. temporal full width half maximum (FWHM). }
    \label{fig:fig1}
\end{figure}

We utilize dispersive Fourier transform (DFT) as the key tool for pulse-wise spectral intensity measurements. DFT maps the optical spectrum of each individual pulse onto a temporally stretched waveform via group-velocity dispersion in a dispersive medium, allowing single-shot spectral acquisition at the repetition rate of the source using a fast photodetector and oscilloscope. Figure~\ref{fig:fig1}a shows the experimental setup used for shot-to-shot supercontinuum characterization. The system consisted of a femtosecond pulsed laser source (Toptica DFC, compensated to 170\,fs, 80 MHz) followed by an erbium-doped fiber amplifier (EDFA) and a programmable WaveShaper used to control the applied chirp before the highly nonlinear fiber (HNLF). The generated spectra were characterized using both dispersive Fourier transform (DFT) measurements and optical spectrum analyzer (OSA) measurements.

Figures~\ref{fig:fig1}b and c show representative supercontinuum spectra generated using chirped excitation pulses with a temporal duration of approximately 700 fs. Figure~\ref{fig:fig1}b shows the reconstructed spectrum obtained from the DFT acquisition, while Fig.~\ref{fig:fig1}c shows the corresponding OSA measurement acquired under the same operating condition. Both measurements exhibit the same spectral features and bandwidth evolution, confirming that the DFT based spectral reconstruction is in good agreement with the OSA measurements.

The applied chirp was used to systematically vary the temporal duration of the excitation pulses prior to supercontinuum generation. As shown in Fig.~\ref{fig:fig1}d, the integrated standard deviation initially exhibited relatively large values at short pulse durations and gradually decreased as the pulse duration increased. This behavior indicates that shorter excitation pulses produced stronger shot-to-shot spectral fluctuations, while longer chirped pulses generated comparatively more stable spectral evolution within the analyzed dynamic range.

Figure~\ref{fig:fig1}e shows the relationship between the applied second-order chirp $\beta_2$ and the measured pulse duration. The blue data points correspond to experimentally measured pulse durations obtained for different applied dispersion values. The green curve represents the theoretical estimation of the pulse duration evolution obtained by fitting the experimental data using the expected temporal broadening induced by the applied $\beta_2$. The agreement between the experimental measurements and the fitted model confirms the controlled chirp-dependent pulse duration sweep used throughout the study.

\subsection*{Simulation Side Statistical Signatures}

\begin{figure}[H]
    \centering
    \includegraphics[width=14cm]{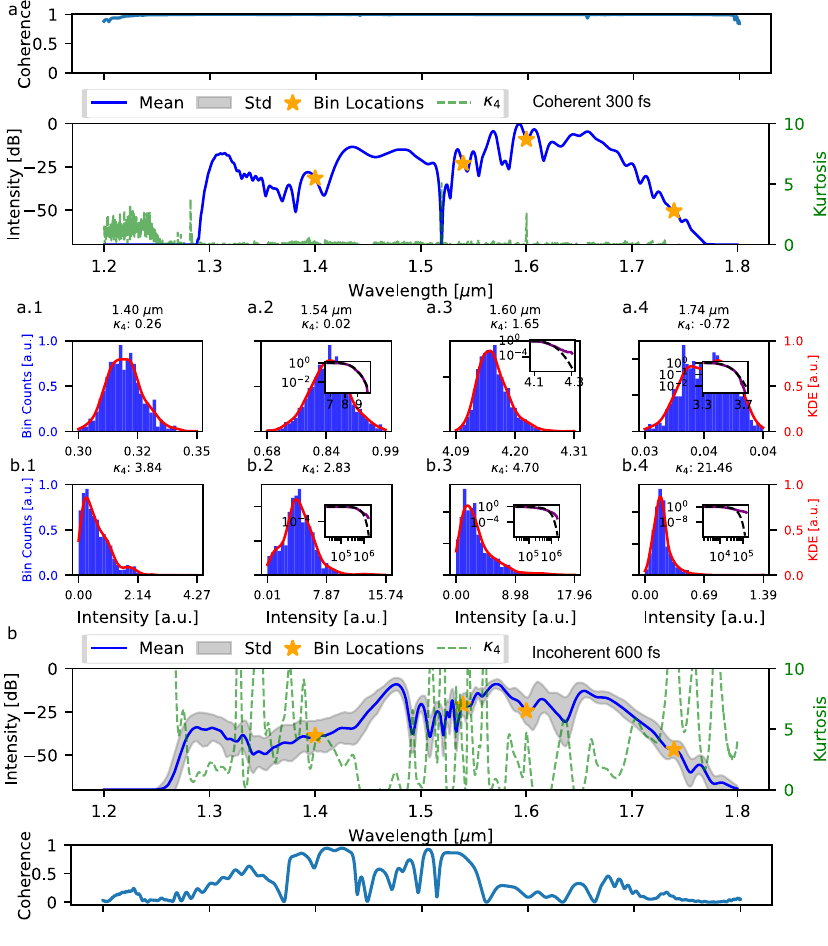}
    \caption{Statistical aspects of supercontiniuum spectra in the coherent and incoherent domain from simulations. a-b) Supercontinuum output spectra generated from (a) 300 fs gaussian pump pulses and (b) 700 fs gaussian pump pulses with respective histograms and kernel density estimation (KDE) in panels (1) to (4) for selected spectral bins at 1400, 1540, 1600, and 1740~nm, each. The insets within the histograms show the corresponding complementary cummuliative distribution function. }
    \label{fig:fig2}
\end{figure}

We employ simulations based on the nonlinear Schödinger equation to create a relation between the well-known first order degree of coherence and our statistical measures. Figure~\ref{fig:fig2} shows representative simulation results for short and long duration excitation pulses together with the corresponding spectral intensity statistics extracted at selected wavelength bins centered at 1400, 1540, 1600, and 1740 nm. For each wavelength bin, the shot-to-shot intensity fluctuations were analyzed using histograms and complementary cumulative distribution functions (CCDFs).

The statistical distributions exhibited clear differences between the coherent 300 fs regime (cf. Fig.~\ref{fig:fig2}a) and the more turbulent 600 fs regime (cf. Fig.~\ref{fig:fig2}b). In the 300 fs case, the spectral intensity distributions remained comparatively closer to Gaussian statistics across most wavelength bins. In contrast, the 600 fs excitation case showed strong deviations from Gaussian behavior, particularly near the spectral edges of the generated supercontinuum.

The strongest deviation was observed at the 1.74 $\mu$m wavelength bin, shown in Fig.~\ref{fig:fig2}a(4) and Fig.~\ref{fig:fig2}b(4). In the 600 fs case, the intensity histogram exhibited a pronounced heavy-tailed distribution together with a substantially larger kurtosis value compared to the 300 fs case. This behavior is further supported by the CCDF analysis shown in the subfigures of Fig.~\ref{fig:fig2}a(4) and Fig.~\ref{fig:fig2}b(4). We further use kernel density estimation (KDE) to estimate the probability density function from sample data by placing a smooth kernel, often a Gaussian bump, at each data point and averaging the result. The dashed curves in the insets of Figs.~\ref{fig:fig2}a(2-4) and b(2-4) represent the Gaussian distribution CCDF corresponding to the measured sample variance, while the magenta curves represent the CCDF extracted from the KDE of the measured distributions. The 600 fs excitation case exhibited a significantly larger probability of high-intensity tail events, indicating intermittent spectral fluctuations and increased stochasticity in the supercontinuum evolution.

A similar behavior was observed on both the shortest and longest wavelengths sides of the generated supercontinuum spectrum, where the kurtosis values fluctuated strongly across wavelength in the intermediate pulse duration regime around 600 fs. These fluctuations indicate that the transition toward turbulent supercontinuum generation is not spectrally uniform but instead emerges preferentially in localized wavelength regions. In contrast, wavelength bins near 1.54 $\mu$m and 1.60 $\mu$m exhibited comparatively smaller deviations from Gaussian statistics and lower tail probabilities.

When we compare the statistical behavior observed in the intensity distributions against the first-order coherence calculated from the simulated optical fields we find that the 300~fs excitation case maintained high spectral coherence across the generated bandwidth, whereas the 600 fs case exhibited significantly reduced coherence values. This reduction in coherence coincided with the appearance of heavy-tailed intensity statistics and enhanced kurtosis, supporting the interpretation that the observed statistical fluctuations originate from increasingly incoherent and turbulent nonlinear dynamics.

\subsection*{Experimental Statistical Signatures}

\begin{figure}[H]
    \centering
    \includegraphics[width=14cm]{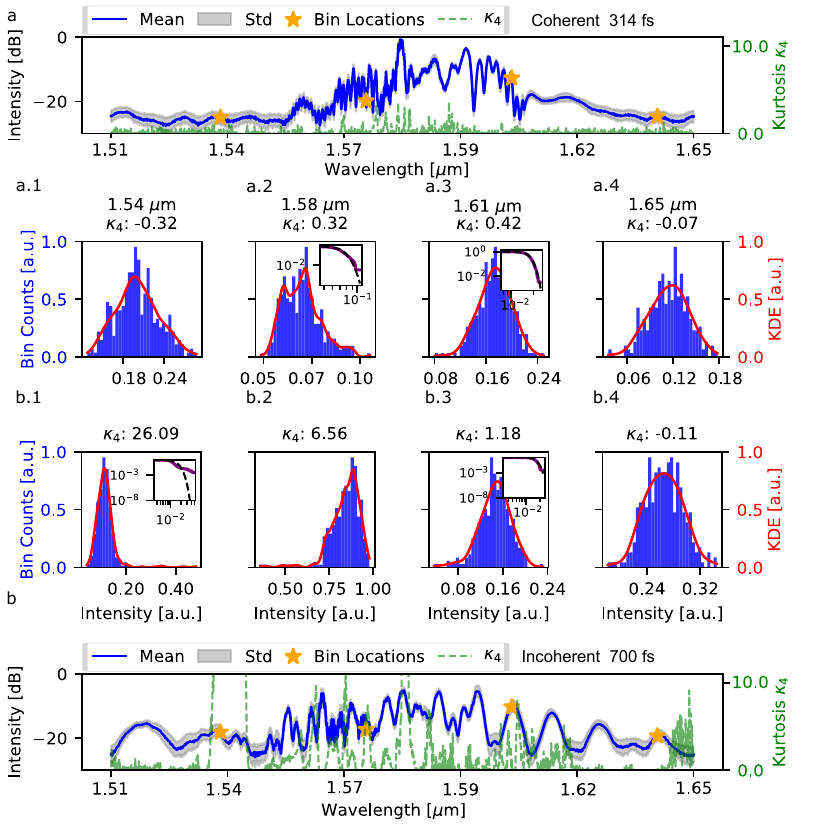}
    \caption{Statistical aspects of supercontiniuum spectra in the coherent and incoherent domain measured with an experimental DFT setup. a-b) Supercontinuum output spectra generated from (a) 314 fs gaussian pump pulses and (b) 700 fs gaussian pump pulses with respective histograms and kernel density estimation (KDE) in panels (1) to (4) for selected spectral bins at 1540, 1580, 1610, and 1650~nm each. The insets within the histograms show the corresponding complementary cummuliative distribution function.}
    \label{fig:fig3}
\end{figure}

Experimentally measured shot-to-shot spectral statistics in Figure~\ref{fig:fig3} obtained from DFT acquisitions reproduce the same qualitative statistical transition as observed in the numerical simulations. Representative wavelength bins centered at 1540, 1580, 1610, and 1650~nm were analyzed for different pulse durations. In the shorter pulse duration regime, e.g. at 314~fs shown in Fig.~\ref{fig:fig3}a, the spectral intensity distributions remained relatively narrow and closer to Gaussian statistics. However, in case of the 700~fs excitation in Fig.~\ref{fig:fig3}b, strong deviations from Gaussian behavior emerged in selected wavelength regions. 

The largest deviation from Gaussian statistics was observed on the short wavelengths side of the DFT spectrum near 1.54~$\mu$m. In this wavelength region, the intensity histograms exhibited broader heavy-tailed distributions together with CCDF curves containing a significantly larger number of high-intensity events compared to the Gaussian reference distribution. The increased kurtosis values observed in these wavelength bins indicate the presence of intermittent extreme fluctuations during the supercontinuum evolution. It is particularly noteworthy that the kurtosis in the electronic noise domains at the edges of the recording domain is close to 0.0. Observations of high kurtosis values hence originate from optical phenomena.

Compared with the simulation results, the experimental data showed stronger wavelength dependent kurtosis fluctuations in the long pulse duration regime. The appearance of spectrally localized heavy-tailed statistics in both the simulations and experiments demonstrates good qualitative agreement between the numerical model and the experimental measurements. These results support the interpretation that the observed fluctuations originate from increasingly incoherent nonlinear spectral dynamics as modulation instabilities become dominant with increasing pulse duration.

\subsection*{Integrated Metrics Across Pulse Duration}

\begin{figure}[H]
    \centering
    \includegraphics[width=14cm]{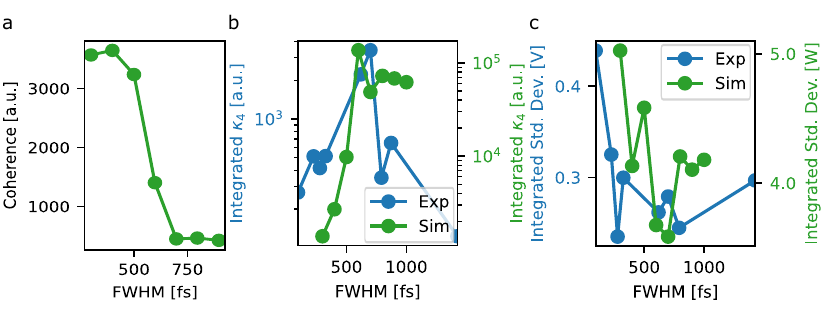}
    \caption{Statistical metrics to distinguish turbulenced supercontinuum generation. a) First-order degree of coherence against pulse duration calculated on simulation data. b) Simulation and experimental result comparsion for integrated kurtosis over pulse duration. c) Simulation and experimental result comparsion for integrated standard deviation over pulse duration.}
    \label{fig:fig4}
\end{figure}

Figure~\ref{fig:fig4} summarizes the evolution of the integrated statistical observables as a function of pulse duration for both the simulations and experiments. Figure~\ref{fig:fig4}a shows the first-order coherence calculated from the simulated optical fields. The coherence decreased drastically between 400 fs and 700 fs pulse duration, indicating a transition from coherent supercontinuum generation driven by soliton fission toward increasingly stochastic spectral evolution of modulation instabilities.

The integrated kurtosis values extracted from both simulation and experimental data are shown in Fig.~\ref{fig:fig4}b. Although the absolute kurtosis values differed between the numerical and experimental measurements, both cases exhibited a similar overall trend. In the simulations, the integrated kurtosis reached a maximum near 600 fs before decreasing and fluctuating at longer pulse durations. In the experimental measurements, the corresponding maximum occurred near 700 fs. While the plateau-like behavior observed in the simulations at larger pulse durations was not fully reproduced experimentally, both datasets consistently showed enhanced kurtosis in the intermediate pulse duration regime associated with reduced coherence.

Figure~\ref{fig:fig4}c shows the integrated standard deviation calculated within the first 20 dB spectral range. In both the simulations and experiments, the integrated standard deviation decreased with increasing pulse duration. This trend indicates that the overall altitude of spectral deviation becomes smaller as the excitation pulses become longer with chirp.

\section*{Discussion}
The statistical behavior observed in the experimental measurements showed strong qualitative agreement with the numerical simulations, supporting the validity of both the generalized nonlinear Schrödinger equation model and the DFT based experimental acquisition system. In both cases, increasing pulse duration resulted in reduced coherence together with stronger heavy-tailed intensity statistics localized at specific wavelength regions of the supercontinuum spectrum. The appearance of enhanced kurtosis and non-Gaussian CCDF behavior in the experimentally measured spectra confirmed that the DFT system was capable of capturing intermittent extreme fluctuations during supercontinuum generation. In particular, the wavelength-dependent kurtosis variations observed in the incoherent excitation regime were reproduced experimentally with similar spectral localization trends as the simulations. Although the integrated kurtosis maximum occurred near 600 fs in the simulations and around 700 fs in the experiments, the overall behavior remained consistent. This discrepancy may originate from uncertainties in the pulse duration measurements obtained using the optical pulse autocorrelator, especially in the low $\beta_2$ region shown in Fig.~\ref{fig:fig1}e where the measured pulse durations fluctuate relative to the theoretical broadening estimation.

The obtained results also highlight both the usefulness and the limitations of the employed statistical metrics. The integrated standard deviation decreased with increasing pulse duration in both the simulations and experiments, indicating an overall transition in the dynamic regime of the generated spectra. However, the standard deviation alone was insufficient to identify the degree of turbulence because its behavior exhibited an opposite trend compared to the spectral coherence. In contrast, the integrated kurtosis exhibited a pronounced maximum in the intermediate pulse-duration regime. Notably, we obsersed a drastic drop in the simulated coherence at the point of maximum integrated kurtosis. This correspondence suggests that kurtosis provides additional sensitivity to intermittent rare events and spectrally localized stochastic behavior. The CCDF analysis further confirmed the presence of heavy-tailed intensity distributions by showing higher probabilities of extreme spectral events relative to Gaussian statistics. 

Nevertheless, the present approach still contains important limitations. The DFT technique measures intensity-only information and does not preserve the complex optical phase, preventing direct experimental evaluation of field-based coherence. Moreover, both the dynamic range of the real-time scope, typically 8 to 12 bits, and the shot noise of the ultrafast detector prohibit the observation of more subtle spectral components of the supercontinuum. In addition, finite oscilloscope bandwidth, spectral reconstruction uncertainties, and limited sampling rates may influence the retrieved probability distributions, particularly in the extreme-event tails.

Yet, overall, the presented statistical framework provides experimentally accessible tools for navigating coherent, turbulent, and incoherent operating regimes in supercontinuum generation. Integrated kurtosis and variance measurements can serve as practical indicators for turbulence screening without requiring direct access to the optical phase. Furthermore, wavelength-resolved kurtosis and CCDF analysis enable localization of extreme-event regions within the generated spectrum. While coherence remains an important benchmark metric, additional nonlinear dynamics indicators, such as Lyapunov exponent analysis and Pearson correlation analysis, could further strengthen the characterization of chaotic supercontinuum evolution. 

Beyond turbulence identification, the observed heavy-tailed and stochastic spectral statistics may also offer opportunities for applications such as physical random-number generation, stochastic optical sampling, and data-driven generative artificial intelligence systems that benefit from complex nonlinear fluctuations. Future work combining intensity measurements with second-order correlation and dynamical analyses may provide a more complete description of turbulence formation in broadband nonlinear fiber systems.

\section*{Methods}
\subsection*{Numerical Model}
Supercontinuum generation in the highly nonlinear fiber (HNLF) was modeled using the generalized nonlinear Schrödinger equation (GNLSE) 
. Pulse propagation was solved using a symmetric split-step Fourier method in which the linear dispersive operator was evaluated in the frequency domain and the nonlinear contribution was integrated using a fourth-order Runge--Kutta (RK4) scheme. The linear propagation step was expressed through the half-step operator $\exp[i\beta(\omega)\delta z/2]$, where $\beta(\omega)$ represents the frequency dependent propagation constant and $\delta z$ is the propagation increment.
The simulated propagation length was $L = 5$ m and the integration employed $N_z = 1000$ longitudinal steps. The temporal simulation window was set to $T = 50$ ps with $N_t = 2^{14}$ discrete temporal samples, corresponding to a temporal resolution of approximately 3 fs. The HNLF parameters used in the simulations were $\beta_2 = 9.3699\times10^{-28}\,\mathrm{s^2\,m^{-1}}$, $\beta_3 = 7.6891\times10^{-42}\,\mathrm{s^3\,m^{-1}}$, $\beta_4 = 3.1340\times10^{-55}\,\mathrm{s^4\,m^{-1}}$, attenuation coefficient $\alpha = 0.1\,\mathrm{m^{-1}}$, nonlinear coefficient $\gamma = 10.8\times10^{-3}\,\mathrm{W^{-1}\,m^{-1}}$, inverse effective mode area $A_{\mathrm{eff}}^{-1}=8\times10^{10}\,\mathrm{m^{-2}}$, and core radius $R = 1.92\times10^{-6}$ m. These parameters were selected to represent an HNLF designed for supercontinuum generation near 1550 nm.

The simulations were initialized using a 300 fs Gaussian input pulse with Gaussian photon-noise seeding applied independently to each spectral mode. Chirp was introduced through controlled variation of $\beta_2$, while the peak power corresponding to the Fourier-limited input pulse was preserved at 3~kW. This approach enabled a systematic investigation of pulse-duration-dependent transitions in supercontinuum dynamics while maintaining comparable nonlinear excitation conditions.

\subsection*{Experimental Setup}
The experimental configuration consisted of a femtosecond laser source (Toptica DFC, 80~MHz repetition rate, 1560~nm center wavelengths, 30~nm spectral bandwidth) followed by an erbium-doped fiber amplifier (Thorlabs EDFA300P) and a programmable WaveShaper (Coherent 1000A/X) used for spectral and chirp control. The conditioned pulses were launched into the HNLF for supercontinuum generation. A fiber-coupled beam splitter (50:50) directed the signal toward the dispersive Fourier transform (DFT) measurement path, while average spectral measurements were independently acquired using an optical spectrum analyzer (OSA).

The DFT readout employed 800 m of dispersion-compensating fiber (DCF) from M2 Optics, Inc., with a dispersion coefficient of 70 ps/nm. The temporally stretched signal was detected using a high-speed photodiode (Albis PMY30A-L, 34.4 GHz bandwidth). The electrical waveform was recorded using a real-time oscilloscope (Keysight UXR1104B, 110 GHz bandwidth, 256 GS/s sampling rate, 10-bit analog-to-digital conversion). Independent spectral validation measurements were performed using an optical spectrum analyzer (Yokogawa AQ6375E) with 0.2 nm spectral resolution over a wavelength range from 1000 nm to 2500 nm. Pulse durations were measured using an APE PulseCheck 50 autocorrelator configured with a 15 ps temporal scan range.

\subsection*{DFT Calibration And Spectral Reconstruction}
We acquired a long oscilloscope trace containing multiple consecutive pulses for each operating condition corresponding to a different applied $\beta_2$. We then identified the maximum of the trace which was used together with the reference signal of the laser source to define the center of the 12.5 ns temporal frames corresponding to the 80~MHz pulse repetition rate of the laser. Each frame was subsequently centered such that the maximum pulse intensity was aligned to the middle of the acquisition window before the full trace was split into stacked single-shot pulse records.

For spectral calibration, a corresponding OSA spectrum was measured under identical operating conditions. The stacked DFT traces were converted to logarithmic scale, after which the far-left and far-right points lying within the first 10 dB above the noise floor were selected in both the DFT and OSA spectra. The maxima of the two traces were aligned after the selected spectral regions were resampled using the SciPy.signal package to ensure identical discrete sampling.

A peak-finding procedure was then applied to construct matched peak points between the temporal DFT trace and the OSA frequency-domain spectrum. These matched points were used to fit a linear polynomial describing the time-to-frequency mapping. The fitted relation was extrapolated over the full temporal window and subsequently converted into wavelength coordinates. When the reconstructed wavelength axis exceeded approximately 178 nm, an additional linear-scale recalibration procedure based on wavelength-to-time matching was applied using the same resampling strategy.

During the initial calibration stage, the reconstructed wavelength axis occasionally extended from approximately 1400 nm to 1870 nm due to extrapolation errors introduced during the time to wavelength fitting procedure. This mismatch was corrected using the recalibration workflow described above, resulting in the corrected wavelength axis used throughout the analysis.

\subsection*{Statistical Analysis}

Shot-to-shot spectra were analyzed at representative wavelength bins selected from the supercontinuum bandwidth. For the numerical simulations, the selected wavelength bins were centered at 1400, 1540, 1600, and 1740 nm. For the experimental measurements, the selected wavelength bins were centered at 1540, 1580, 1610, and 1650 nm. These wavelength regions were chosen to compare fluctuations across different spectral regions of the generated supercontinuum.

For each wavelength bin, the shot-to-shot intensity distribution was analyzed using histograms and complementary cumulative distribution functions (CCDFs). The CCDF of a random intensity variable $I$ was defined as
\begin{equation}
\mathrm{CCDF}(I)=P(X>I)=1-F(I),
\label{eq:CCDF}    
\end{equation}
where $F(I)$ is the cumulative distribution function of the measured intensity distribution. The CCDF representation emphasizes the probability of rare high-intensity events and therefore provides a sensitive indicator of heavy-tailed statistics. Deviations of the measured CCDF from the Gaussian-distribution CCDF indicate the emergence of non-Gaussian fluctuations associated with turbulent or chaotic supercontinuum dynamics.

To quantify the skewness of the distribution tails, the kurtosis of the spectral intensity distribution was calculated as
\begin{equation}
\kappa_4 =\frac{\langle (I-\mu)^4 \rangle}{\sigma^4},
\label{eq: kurtosis}
\end{equation}
where $\mu$ is the mean intensity and $\sigma$ is the standard deviation of the distribution. Larger kurtosis values correspond to distributions containing stronger rare-event contributions and heavier tails compared to Gaussian statistics. Therefore, kurtosis provides a compact statistical metric for identifying spectral regions exhibiting intermittent or turbulent fluctuations.

To compare operating regimes across the full spectral bandwidth, wavelength-integrated statistical observables were calculated. Integrated kurtosis was used to quantify the emergence of heavy-tailed intensity distributions associated with rare, high-amplitude fluctuations, while integrated variance and standard deviation were used to characterize the overall fluctuation strength of the generated spectra.

First-order spectral coherence was evaluated only for the numerical simulations because the experimental DFT measurements provide intensity-only information and do not preserve the complex optical field. The first-order coherence was calculated as
\begin{equation}
|g^{(1)}_{12}(\lambda)|=
\left|
\frac{\langle E_1^{*}(\lambda)E_2(\lambda)\rangle}
{\sqrt{\langle |E_1(\lambda)|^2\rangle \langle |E_2(\lambda)|^2\rangle}}
\right|
\label{eq:coherence}
\end{equation}
where $E_1(\lambda)$ and $E_2(\lambda)$ are independently generated spectral fields at wavelength $\lambda$. This metric quantifies the shot-to-shot phase stability of the supercontinuum and is widely used as a reference indicator for distinguishing coherent and chaotic spectral evolution. Reduced coherence values correspond to increasingly stochastic spectral dynamics and therefore provide a benchmark against which the experimentally accessible intensity-based statistical observables can be compared.

\section*{Acknowledgements}
We also acknowledge Julian Hniopek and his team at Leibniz-IPHT for providing the computing infrastructure needed for this project. We further acknowledge Keysight Germany for access to the real-time scope.

\section*{Data availability}
Data is available on request.

\bibliographystyle{ieeetr}
\bibliography{statistical_metrics_references}

@article{perego_complexity_2022,
	title = {Complexity of modulation instability},
	volume = {4},
	issn = {2643-1564},
	url = {https://link.aps.org/doi/10.1103/PhysRevResearch.4.L022057},
	doi = {10.1103/PhysRevResearch.4.L022057},
	pages = {L022057},
	number = {2},
	journal = {Phys. Rev. Research},
	author = {Perego, Auro M. and Bessin, Florent and Mussot, Arnaud},
	urldate = {2024-10-08},
	date = {2022-06-13},
	year = {2022},
	langid = {english},
}

@article{solli_optical_2007,
	title = {Optical rogue waves},
	volume = {450},
	rights = {http://www.springer.com/tdm},
	issn = {0028-0836, 1476-4687},
	url = {https://www.nature.com/articles/nature06402},
	doi = {10.1038/nature06402},
	pages = {1054--1057},
	number = {7172},
	journal = {Nature},
	author = {Solli, D. R. and Ropers, C. and Koonath, P. and Jalali, B.},
	urldate = {2024-10-08},
	date = {2007-12},
	year = {2007},
	langid = {english},
}

@article{erkintalo_statistical_2010,
	title = {On the statistical interpretation of optical rogue waves},
	volume = {185},
	rights = {http://www.springer.com/tdm},
	issn = {1951-6355, 1951-6401},
	url = {http://link.springer.com/10.1140/epjst/e2010-01244-9},
	doi = {10.1140/epjst/e2010-01244-9},
	pages = {135--144},
	number = {1},
	journal = {Eur. Phys. J. Spec. Top.},
	author = {Erkintalo, M. and Genty, G. and Dudley, J.M.},
	urldate = {2024-10-08},
	date = {2010-07},
	year = {2010},
	langid = {english},
}

@misc{borlaug_extreme_2008,
	title = {Extreme value statistics and the Pareto distribution in silicon photonics},
	url = {http://arxiv.org/abs/0809.2565},
	number = {{arXiv}:0809.2565},
	publisher = {{arXiv}},
	author = {Borlaug, David and Fathpour, Sasan and Jalali, Bahram},
	urldate = {2024-10-06},
	date = {2008-09-15},
	year = {2008},
	langid = {english},
	eprinttype = {arxiv},
	eprint = {0809.2565 [physics]},
}

@article{dudley_harnessing_2008,
	title = {Harnessing and control of optical rogue waves in supercontinuum generation},
	volume = {16},
	rights = {https://doi.org/10.1364/{OA}\_License\_v1\#{VOR}-{OA}},
	issn = {1094-4087},
	url = {https://opg.optica.org/oe/abstract.cfm?uri=oe-16-6-3644},
	doi = {10.1364/OE.16.003644},
	pages = {3644},
	number = {6},
	journal = {Opt. Express},
	author = {Dudley, John M. and Genty, Goëry and Eggleton, Benjamin J.},
	urldate = {2024-10-06},
	date = {2008-03-17},
	year = {2008},
	langid = {english},
}

@article{narhi_machine_2018,
	title = {Machine learning analysis of extreme events in optical fibre modulation instability},
	volume = {9},
	issn = {2041-1723},
	url = {https://www.nature.com/articles/s41467-018-07355-y},
	doi = {10.1038/s41467-018-07355-y},
	pages = {4923},
	number = {1},
	journal = {Nat Commun},
	author = {Närhi, Mikko and Salmela, Lauri and Toivonen, Juha and Billet, Cyril and Dudley, John M. and Genty, Goëry},
	urldate = {2024-10-06},
	date = {2018-11-22},
	year = {2018},
	langid = {english},
}

@article{zakharov_modulation_2009,
	title = {Modulation instability: The beginning},
	volume = {238},
	rights = {https://www.elsevier.com/tdm/userlicense/1.0/},
	issn = {01672789},
	url = {https://linkinghub.elsevier.com/retrieve/pii/S0167278908004223},
	doi = {10.1016/j.physd.2008.12.002},
	shorttitle = {Modulation instability},
	pages = {540--548},
	number = {5},
	journal = {Physica D: Nonlinear Phenomena},
	author = {Zakharov, V.E. and Ostrovsky, L.A.},
	urldate = {2024-10-06},
	date = {2009-03},
	year = {2009},
	langid = {english},
}

@article{hammani_spectral_2011,
	title = {Spectral dynamics of modulation instability described using Akhmediev breather theory},
	volume = {36},
	rights = {https://doi.org/10.1364/{OA}\_License\_v1\#{VOR}},
	issn = {0146-9592, 1539-4794},
	url = {https://opg.optica.org/abstract.cfm?URI=ol-36-11-2140},
	doi = {10.1364/OL.36.002140},
	pages = {2140},
	number = {11},
	journal = {Opt. Lett.},
	author = {Hammani, K. and Wetzel, B. and Kibler, B. and Fatome, J. and Finot, C. and Millot, G. and Akhmediev, N. and Dudley, J. M.},
	urldate = {2024-10-06},
	date = {2011-06-01},
	year = {2011},
	langid = {english},
}

@article{dudley_instabilities_2014,
	title = {Instabilities, breathers and rogue waves in optics},
	volume = {8},
	issn = {1749-4885, 1749-4893},
	url = {https://www.nature.com/articles/nphoton.2014.220},
	doi = {10.1038/nphoton.2014.220},
	pages = {755--764},
	number = {10},
	journal = {Nature Photon},
	author = {Dudley, John M. and Dias, Frédéric and Erkintalo, Miro and Genty, Goëry},
	urldate = {2024-10-06},
	date = {2014-10},
	year = {2014},
	langid = {english},
}

@article{akhmediev_extreme_2009,
	title = {Extreme waves that appear from nowhere: On the nature of rogue waves},
	volume = {373},
	rights = {https://www.elsevier.com/tdm/userlicense/1.0/},
	issn = {03759601},
	url = {https://linkinghub.elsevier.com/retrieve/pii/S0375960109004939},
	doi = {10.1016/j.physleta.2009.04.023},
	shorttitle = {Extreme waves that appear from nowhere},
	pages = {2137--2145},
	number = {25},
	journal = {Physics Letters A},
	author = {Akhmediev, N. and Soto-Crespo, J.M. and Ankiewicz, A.},
	urldate = {2024-10-06},
	date = {2009-06},
	year = {2009},
	langid = {english},
}

@article{kraych_statistical_2019,
	title = {Statistical Properties of the Nonlinear Stage of Modulation Instability in Fiber Optics},
	volume = {123},
	issn = {0031-9007, 1079-7114},
	url = {https://link.aps.org/doi/10.1103/PhysRevLett.123.093902},
	doi = {10.1103/PhysRevLett.123.093902},
	pages = {093902},
	number = {9},
	journal = {Phys. Rev. Lett.},
	author = {Kraych, Adrien E. and Agafontsev, Dmitry and Randoux, Stéphane and Suret, Pierre},
	urldate = {2024-10-06},
	date = {2019-08-28},
	year = {2019},
	langid = {english},
}

@article{wetzel_real-time_2012,
	title = {Real-time full bandwidth measurement of spectral noise in supercontinuum generation},
	volume = {2},
	issn = {2045-2322},
	url = {https://www.nature.com/articles/srep00882},
	doi = {10.1038/srep00882},
	pages = {882},
	number = {1},
	journal = {Sci Rep},
	author = {Wetzel, B. and Stefani, A. and Larger, L. and Lacourt, P. A. and Merolla, J. M. and Sylvestre, T. and Kudlinski, A. and Mussot, A. and Genty, G. and Dias, F. and Dudley, J. M.},
	urldate = {2024-10-06},
	date = {2012-11-28},
	year = {2012},
	langid = {english},
}

\section*{Acknowledgements}
Funding. Carl-Zeiss Stiftung, Nexus program (P2021-05-025).
We acknowledge that this work was made possible by funding from the Carl Zeiss
Foundation through the NEXUS program (project P2021-05-025)



\end{document}